\documentclass[prd,superscriptaddress,amsfonts,amssymb,amsmath,showpacs]{revtex4-2}
\usepackage{bm}
\usepackage{amsfonts}
\usepackage{latexsym}
\usepackage{graphicx}
\usepackage{amsmath}
\usepackage{palatino}
\usepackage{mathpazo}
\usepackage{textcomp}
\usepackage{float}
\usepackage{booktabs}
\usepackage{dcolumn}
\usepackage{booktabs}
\usepackage{multirow}
\usepackage{hyperref}
\hypersetup{colorlinks,citecolor=blue}
\usepackage{amsmath}
\usepackage{xcolor}
\usepackage{orcidlink}
\usepackage{subcaption}
\usepackage{commath}
\def\jnl@style{\it}
\def\aaref@jnl#1{{\jnl@style#1}}
\def\aaref@jnl#1{{\jnl@style#1}}

\allowdisplaybreaks[1]
 \allowdisplaybreaks
\begin{document}
\title{Role of Complexity on the Minimal Deformation of Black Holes}

\author{Z. Yousaf}
 \email[Email: ]{zeeshan.math@pu.edu.pk}
\affiliation{Department of Mathematics, University of the Punjab, Quaid-i-Azam Campus, Lahore-54590, Pakistan.}

\author{Kazuharu Bamba}
 \email[Email: ]{bamba@sss.fukushima-u.ac.jp}
\altaffiliation[Corresponding author ]{}
\affiliation{Faculty of Symbiotic Systems Science,
Fukushima University, Fukushima 960-1296, Japan.}%

\author{Bander Almutairi}
\email[Email: ]{baalmutairi@ksu.edu.sa}
\affiliation{Department of Mathematics, College of Science,\\
King Saud University, P.O.Box 2455 Riyadh 11451, Saudi Arabia.}%

\author{S. Khan}
\email[Email: ]{suraj.pu.edu.pk@gmail.com}
\affiliation{Department of Mathematics, University of the Punjab, Quaid-i-Azam Campus, Lahore-54590, Pakistan.}

\author{M. Z. Bhatti}
\email[Email: ]{mzaeem.math@pu.edu.pk}
\affiliation{Department of Mathematics, University of the Punjab, Quaid-i-Azam Campus, Lahore-54590, Pakistan.}

%\author{Z. Yousaf$^1$ \thanks{zeeshan.math@pu.edu.pk} Kazuharu Bamba$^2*$
%\thanks{bamba@sss.fukushima-u.ac.jp; Corresponding author}, Bander Almutairi$^3$ \thanks{baalmutairi@ksu.edu.sa}, S.
%Khan$^1$ \thanks{suraj.pu.edu.pk@gmail.com}, and M. Z. Bhatti$^1$
%\thanks{mzaeem.math@pu.edu.pk}\\
%$^1$ Department of Mathematics, University of the Punjab,\\
%Quaid-i-Azam Campus, Lahore-54590, Pakistan.\\
%$^2$  Faculty of Symbiotic Systems Science,\\
%Fukushima University, Fukushima 960-1296, Japan.\\
%$^3$ Department of Mathematics, College of Science,\\
%King Saud University, P.O.Box 2455 Riyadh 11451, Saudi Arabia.}


\begin{abstract}

We investigate spherically symmetric classes of anisotropic
solutions within the realm of a schematic gravitational decoupling scheme, primarily decoupling through minimal geometric
deformation, applied to non-rotating, ultra-compact,
self-gravitational fluid distributions. In this respect, we employ
the minimal complexity factor scheme
to generate physically realistic models for
anisotropic matter distributions, using a well-behaved model. The zero-complexity factor condition enables us to
determine the deformation function for solving the decoupled system.
We explore all the structure-defining scalar variables, such as
density inhomogeneity, strong energy
condition, density homogeneity, and the complexity factor (an alloy of
density inhomogeneity and pressure anisotropy) for the decoupling
constant ranging between $0$ and $1$. We observe that the anisotropy vanishes when the coupling constant is set to unity. This finding holds significance as it implies that, in the context of a zero-complexity factor approach, an anisotropic matter distribution becomes perfect without requiring any isotropy requirements. This work effectively explored the impact of complexity on the composition of self-gravitational stellar distributions. This effective approach enables the development of new, physically realistic isotropic stellar models for anisotropic matter distributions. Additionally, our findings indicate that the complexity factor in static, spherically symmetric self-gravitational objects can significantly affect the nature of the matter distribution within these systems. It is concluded that the minimally deformed Durgapal-IV model features an increasing pressure profile, and the local anisotropy of pressure vanishes throughout the model under complexity-free conditions.
\end{abstract}
\maketitle

\section{Introduction}

The Einstein gravitational-field equations (EFEs), which describe
the effects of spacetime curvature, present significant challenges
due to their complex and nonlinear nature when attempting to
construct closed-form analytical solutions. In this regard, Schwarzschild made a groundbreaking attempt by constructing an interior solution to the EFEs, endowed with spherical symmetry and isotropic matter content \cite{schwarzschild1916gravityfield}. He also solved
the EFEs for the exterior vacuum case with a perfect fluid
configuration. The recognition of Schwarzschild's interior solution
as a three-sphere geometry is attributed to Weyl
\cite{weyl1919statischen}. In subsequent decades, significant
development occurred in the advancement of closed-form analytical
solutions to EFEs. Understanding the structure and evolution of compact stars requires an equation of state (EoS) for dense nuclear matter. This EoS should describe both the crust and core of a neutron star using a single, consistent physical model. Tolman \cite{tolman1939static} made significant
contributions by constructing the internal solutions of the EFEs
within the background of perfect fluid spheres governed by
polytropic EoS. Subsequently, the properties of
polytropes in cosmological spacetimes were explored in detail by
various authors (see
\cite{stuchlik2016general,novotny2017polytropic} for details). The
fascinating characteristics of these spherically symmetric
polytropes have been thoroughly investigated in subsequent studies
\cite{hod2018lower,hod2018analytic,stuchlik2017gravitational}.
Besides proposing an EoS to correlate the matter variables, we can
employ the geometrical constraints on the functions. By utilizing
this scheme, Tolman formulated a family of eight isotropic
solutions. Tolman's closed-form solutions characterize the
configuration of static, stellar structures with a perfect fluid
EoS, where radial and tangential stresses are equal, denoted as
$P_{r} = P_{\theta}$.

However, follow-up studies revealed that spherical symmetry does not
necessarily ensure isotropic pressure. Recognizing this limitation,
the introduction of anisotropic matter ($P_{r} \neq P_{\theta}$)
became crucial for understanding highly dense objects. Lema\^{i}tre \cite{lemaitre1933univers} explored the interiors of stars by solving the equations of general relativity (GR) for matter distributions with anisotropic properties. This approach has opened new avenues of research, attracting scientists interested in understanding the structure and behavior of stellar matter. Ruderman's work \cite{ruderman1972pulsars} highlights
the possibility of anisotropic nuclear matter at ultra-high
densities ($>10^{15} \textmd{g}/\textmd{cm}^{3}$). This implies a
non-isotropic EoS in these regimes. Additionally, Bowers and Liang
\cite{bowers1974anisotropic} elaborated on the importance of
anisotropic EoS for highly dense self-gravitational configurations
of fluid spheres. Finding physically viable solutions to the EFEs in
the presence of anisotropy remains a significant challenge
\cite{delgaty1998physical}. Despite its considerable challenge, a
wide range of works in this direction have been available in the
literature
\cite{herrera1992cracking,herrera1997local,mak2003anisotropic,ivanov2002static,schunck2003general,usov2004electric,deb2018anisotropic,
rahaman2010singularity,varela2010charged,yousaf2022analysis,yousaf2022stability}.
These investigations have successfully addressed the impact of
anisotropy for understanding the evolution and formation of highly
dense stellar distributions. Harko pointed out that
various mechanisms can induce anisotropy in ultra-compact stellar
distributions, including the presence of a type 3A superfluid
\cite{kippenhahn1990stellar}, the existence of a solid core
\cite{ruderman1972pulsars}, pion condensation
\cite{sawyer1972condensed}, or different types of phase transitions
\cite{hartle1975pion}.

The existence of anisotropy gives rise to several characteristics in the matter distribution of self-gravitational objects. For example, a positive anisotropic factor $\Delta\equiv P_{r}-P_{\theta}>0$ introduces a repulsive force (attractive when the anionic factor $\Delta$ is negative) that counters the gravitational collapse. This enables us to construct highly dense structures
with anisotropic fluids compared to isotropic fluids \cite{mak2003anisotropic,ivanov2002static}.
Additionally, positive anisotropy enhances the stability and equilibrium of the self-gravitational distribution.

For decades, self-gravitational compact entities like stars and
galaxies have been crucial tools in unraveling the mysteries of our
vast universe. Large-scale surveys like the Sloan Digital Sky Survey
and the Large Synoptic Survey Telescope provide invaluable data on
these cosmic bodies. However, these objects are incredibly complex
heavenly configurations. Even slight fluctuations within them can
significantly alter their physical properties, potentially offering
new insights into the universe's origin and evolution.
Within this domain, a notable challenge resides in precisely gauging the complex interplay among various physical phenomena transpiring within stars and analogous self-gravitational compact structures.
This is where the notion of a "complexity factor" stems and assumes
paramount importance. This factor seeks to enfold the collective
effects of numerous internal mechanisms and their collective
influence on the system's global dynamics. Within stable
self-gravitational distributions, complexity emerges from the
interconnection of these diverse elements, often giving rise to
emergent phenomena that cannot be easily foreseen based solely on
the behavior of individual components. Presently, ongoing research
across diverse scientific fields is centered on establishing precise
definitions and characterizations of complexity. Within the domain
of compact configurations, especially those of significant mass, the
analysis of complexity frequently entails evaluating metrics related
to information and entropy present within the configuration of the
system. The investigation of ultra-compact stellar configurations
with anisotropic matter content often involves the examination of
the complexity of self-gravitational structures. In physics,
isolated gases are regarded as intricate systems due to their
disorder and high information content, while perfect crystals, with
their periodic behavior and symmetric dispersion, are seen as
possessing minimal complexity. "disequilibrium" as a measure of
complexity in systems.  This measure quantifies the deviation from
the system's most probable, equally likely state, essentially
capturing the system's "distance" from equilibrium.  They argued
that for ideal gases and perfect crystals, the concept of complexity
vanishes when considering both disequilibrium and information
content. Considering the limitations in existing complexity measures
for self-gravitating systems, Herrera \cite{herrera2018new} proposed
a novel approach based on fluid properties. This includes
fundamental quantities like energy density and pressure, capturing
the system's overall content. The complexity is then constructed
using a ``complexity factor", one of the structural scalars derived
from the orthogonal decomposition of the intrinsic curvature.

Building upon their prior work, Herrera \emph{et al.}
\cite{herrera2018definition} extended the complexity concept to
dissipative fluids.  They went beyond mere complexity analysis and
established conditions for evolutionary paths with minimal
complexity. Their investigation revealed the existence of multiple
solutions, with the fluid exhibiting shearing and geodesic
properties in the dissipative regime. Herrera \emph{et al.} further
explored the interplay between complexity and geometry
\cite{herrera2019complexity}. Utilizing axially symmetric
geometries, they identified three distinct complexity categories and
demonstrated the relationship between complexity and symmetry.
Notably, they obtained analytical solutions in this specific case.
On the other hand \cite{herrera2020quasi}, they investigated the
evolution of non-static geometries endowed with spherical symmetry
using the notion of complexity, considering both dissipative and
non-dissipative scenarios. In this context, they constructed
anisotropic stellar-type models and explored the implications for
understanding their evolution using the quasi-homologous condition.
Many researchers have employed this approach to explore the
evolution of stellar distributions through alternative gravitational
models
\cite{bhatti2021electromagnetic,bhatti2021role,yousaf2022f,yousaf2024modeling}.
Herrera et al. \cite{herrera2021hyperbolically} extended the
complexity factor concept to geometries with hyperbolic symmetry.
They incorporated both the Misner-Sharp mass and Tolman mass, along
with various structural scalars obtained through the orthogonal
decomposition of the intrinsic curvature. Interestingly, their
analysis revealed that the Tolman mass exhibits a negative nature in
this specific case. For more related references, see \cite{yousaf2023quasi} (recent review) and references therein.

The minimal geometric deformation (MGD) method has gained attraction
as a powerful and versatile tool for extending isotropic solutions
to EFEs. Its primary advantage lies in its direct and analytical
methodology, which necessitates minimal geometric alterations. This
feature renders it a valuable substitute for current techniques.
Within the context of GR, the gravitational decoupling (GD) facilitated by the minimal
geometric deformation (MGD) approach offers a seamless means of
introducing the necessary alterations to equations or solutions. The MGD
the technique presents a strong capability: providing a pre-existing
solution to EFEs, it enables the discovery of innovative solutions
that integrate an additional gravitational-field source. Primarily,
proposed for the Randall-Sundrum braneworld
\cite{casadio2015minimal,ovalle2016extending,ovalle2009nonuniform,ovalle2010schwarzschild,casadio2012brane,
ovalle2013tolman,casadio2014black,ovalle2015brane}, the MGD scheme
has found significant applications beyond its initial domain.
Significantly, it has played a crucial role in investigating novel
categories of black hole
solutions \cite{casadio2015minimal,ovalle2016extending}. This feat is
based on the application of MGD decoupling to the conventional
Schwarzschild metric, showcasing the method's adaptability in
tackling various gravitational challenges. The original applications
of this scheme also focused on black hole acoustics
\cite{da2017black} and GUP Hawking fermions
\cite{casadio2018generalised}. After its development, the MGD scheme
was employed within the framework of GR
\cite{ovalle2017decoupling,ovalle2018anisotropic}. This
implementation extended the method's capability beyond perfect
fluids, with a key achievement being the generalization of solutions
from isotropic fluids to encompass anisotropic domains. As we will
discuss in the following section, this methodology comprises two
primary components: (i) two independent gravitational sectors,
$T_{\mu\nu}$ and $\mathbb{X}_{\mu\nu}$, which interact solely
gravitationally, and (ii) an MGD scheme introduced in the radial
metric potential $\textsl{g}_{rr}$. This deformation enables the
decoupling of the system into two sets of equations, one for each
source. In the MGD scheme, the source $T_{\mu\nu}$ often serves as the
source for a well-established isotropic interior solution and
$\mathbb{X}_{\mu\nu}$ introduces a local anisotropic behavior into
the system. Consequently, GD is known for extending isotropic
solutions to encompass anisotropic scenarios. Utilizing the MGD
scheme for anisotropic fluids within GR poses a distinctive
challenge. The geometric deformation and the components of the
$\mathbb{X}$-gravitational field source $\mathbb{X}_{\mu\nu}$ remain
undetermined. To tackle this issue, researchers have investigated
diverse approaches to impose additional constraints that facilitate
solving for these unknowns. For example (i) mimic constraints scheme
\cite{da2017black,khan2024complexity}
(ii) some anisotropy mechanism \cite{abellan2020regularity} (iii)
zero complexity factor condition
\cite{albalahi2024electromagnetic,albalahi2024isotropization}.

This work explores anisotropic Class I stellar-type configurations
by constructing solutions to EFEs that adhere to the complexity-free
condition. In this context, we employ one of the effective
theoretical techniques for constructing EFEs solutions: the GD
scheme, which admits the zero-complexity factor condition. The
well-behaved and viable metric ansatz, combined with the anisotropic fluid
sphere is utilized to solve the system of equations. Karmarkar (see \cite{tello2020class} and references therein) introduced the embedding Class I condition, which serves as a versatile and straightforward auxiliary condition for solving the EFEs. This condition is typically applicable to any n-dimensional (pseudo)-Riemannian metric. The Durgapal-IV metric ansatz \cite{durgapal1985analytic} is a relativistic model proposed by M.C. Durgapal. He observed that by considering a simple relation $e^{\phi_{1}}\propto(1+r^{4})^{n}$ (where $n$ denotes a positive real constant), closed-form analytical solutions of the EFEs could be derived.
Based on our findings, the pressure anisotropy increases as the radial variable rises, resulting in a more effective stabilization of the shell compared to the central core areas. Additionally, we observe that contributions from MGD decoupling, characterized by a complexity-free condition, appear to diminish the anisotropy of the stellar structures.

The structure of this paper is as follows: In Sec. \textbf{II}, we briefly review
the formalism of GD primarily in the context of anisotropically
structured spherical matter distributions. Section \textbf{III}
covers the introduction of MGD decoupling along with the formulation
of junction conditions. In Sec. \textbf{IV}, we discuss the
zero-complexity factor constraint and the embedding Class I scheme
for constructing the MGD-based, anisotropic Durgapal-IV
self-gravitational compact stars. Section \textbf{V} presents a
detailed physical analysis of the results, and the concluding
remarks are provided in Sec. \textbf{VI}.

\section{The Standard Protocol for Decoupling Gravitational Sources}

To explore the physical properties inherent to decoupled, self-gravitating compact stars, the modified version of standard gravitational action reads
\cite{ovalle2017decoupling,ovalle2018anisotropic}
\begin{align}\label{x1}
S=\int d^{4}x\sqrt{|\textsl{g}|}\left(\frac{\mathrm{R}}{16\pi}+{L}_{m}
\right)+\beta(corrections),
\end{align}
where $\beta$ corresponds to the new gravitational source. The above action can also be defined as
\begin{align}\label{s1}
S^{\circ}=\int d^{4}x\sqrt{|\textsl{g}|}\left(\frac{\mathrm{R}}{16\pi}+{L}_{m}
+\beta{L}_{\mathbb{X}}\right),
\end{align}
where $\mathrm{R}=\textsl{g}^{\mu\nu}R_{\mu\nu}$ denotes the Ricci scalar, while $R_{\mu\nu}$ and $\textsl{g}^{\mu\nu}$ represent the Ricci tensor and metric tensor, respectively.

Furthermore, the Lagrangian for ordinary matter is ${L}_{m}$, whereas the contribution from the additional gravitational field source, originating from scalar, vector, or tensor fields (beyond the purview of GR), is denoted by  ${L}_{\Phi}$. Additionally, $\textsl{g}$ is the trace of $\textsl{g}^{\mu\nu}$ and $\beta$ is a free parameter that determines the strength of decoupling.
Thus, the EFEs featuring the decoupled stellar fluid configurations can be obtained by varying the action $S^{\circ}$ with respect to $\textsl{g}_{\mu\nu}$, yielding
\begin{align}\label{s2}
G_{\mu\nu}\equiv R_{\mu\nu}-\frac{1}{2}\mathrm{R}\textsl{g}_{\mu\nu}= 8\pi T^{\ast}_{\mu\nu},
\end{align}
where the stress-energy tensor (SET) $T^{\ast}_{\mu\eta}$ is defined as
\begin{align}\label{s3}
T^{\ast}_{\mu\nu}=T_{\mu\nu}+\alpha \mathbb{X}_{\mu\nu}.
\end{align}
Here, $\mathbb{X}_{\mu\nu}$ represents generic type of fluid
distribution ($\mathbb{X}$-gravitational sector, hereafter), induced
by GD. Furthermore, $T^{\ast}_{\mu\nu}$ denotes the interior matter
content of the decoupled, self-gravitational spherical object, which
is approximated with an anisotropic distribution of energy density
($\sigma$), tangential pressure ($P_{\theta}$) and radial pressure
$P_{r}$. Thus, the SET $T^{\ast}_{\mu\nu}$ in canonical form is
defined as
\begin{align}\label{s4}
[T^{\ast}]{^{\mu}_{~\nu}}=(\sigma+P_{\theta})U^{\mu}U_{\nu}-P_{\theta}\delta^{\mu}_{~\nu}+
(P_{r}-P_{\theta})\mathcal{X}^{\mu}\mathcal{X}_{\nu},
\end{align}
where $U^{\mu}$ is the four-velocity and $\mathcal{X}^{\mu}$ denotes the unit four vector along the radial radiation direction and satisfy
\begin{align}\label{s5}
U^{\mu}=(e^{-\phi_{1}/2},0,0,0), \quad \mathcal{X}^{\mu}=(0,e^{-\phi_{2}/2},0,0),
\end{align}
respectively, such that $U^{\mu}U_{\mu}=-\mathcal{X}^{\mu}\mathcal{X}_{\mu}=1$.
In standard coordinates, $x^{\mu}=x^{0,1,2,3}=\{t,r,\vartheta,\varphi\}$, the most general form of the metric with static, spherical symmetry can be expressed as
\begin{align}\label{s6}
ds^{2}=e^{\phi_{1}}dt^{2}-e^{\phi_{2}}dr^{2}-r^{2}d\Omega^{2},
\end{align}
where $d\Omega^{2}=dr^{2}-\sin^{2}\vartheta d\varphi^{2}$ denotes the metric of the unit 2-sphere, while $\phi_{1}=\phi_{1}(r)$ and  $\phi_{2}=\phi_{2}(r)$ are functions of the areal radius $r$ only.
Then, under the spherical symmetry endowed metric ansatz \eqref{s6}, the non-null components of $T^{\ast}_{\mu\nu}$ read
\begin{align}\label{s7}
[T^{\ast}]^{0}_{~0}=\sigma^{\ast}, \quad [T^{\ast}]^{1}_{~1}=-P_{r}^{\ast}, \quad [T^{\ast}]^{2}_{~2}=[T^{\ast}]^{3}_{~3}=-P_{\theta}^{\ast}.
\end{align}
As a consequence, the $tt$, $rr$ and $\vartheta\vartheta$ components of the decoupled EFEs turn out to be
\begin{align}\label{s8}
G^{0}_{~0}&=8\pi [T^{\ast}]^{0}_{~0}:\frac{1}{r^{2}}-\left(\frac{1}{r^{2}}-\frac{\phi_{2}'}{r}\right)e^{\phi_{2}}=8\pi(T^{0}_{~0}+\beta \mathbb{X}^{0}_{~0}),
\\\label{s9}
G^{1}_{~1}&=8\pi [T^{\ast}]^{1}_{~1}:-\frac{1}{r^{2}}+\left(\frac{1}{r^{2}}-\frac{\phi_{2}'}{r}\right)e^{\phi_{2}}=8\pi(T^{1}_{~1}+
\beta\mathbb{X}^{1}_{~1}),
\\\label{s10}
G^{2}_{~2}&=8\pi [T^{\ast}]^{2}_{~2}:\frac{e^{-\phi_{2}}}{4}\left(2\frac{\phi_{2}'}{r}-2\frac{\phi_{1}'}{r}+\phi_{2}'\phi_{1}'
-\phi'^{2}-2\phi_{1}''\right)=8\pi(T^{2}_{~2}+\beta\mathbb{X}^{2}_{~2}),
\end{align}
where $f'\equiv\partial_{r}f$. Because $G^{\mu}_{~\nu}$ is divergence-free, the SET $[T^{\ast}]_{\mu\nu}$ follows the subsequent conservation expression
\begin{align}\label{s11}
0\equiv[T^{\ast}]^{\mu\nu}_{~~~;\nu}=\frac{d[T^{\ast}]^{\mu\nu}}{dx^{\nu}}+\Gamma^{\mu}_{\nu\epsilon}[T^{\ast}]^{\epsilon\nu}
+\Gamma^{\nu}_{\nu\epsilon}[T^{\ast}]^{\mu\epsilon},
\end{align}
which, for $\mu=1$, provides a decoupled hydrostatic equilibrium equation, defined as
\begin{align}\label{s12}
\frac{d}{dr}\left([T^{\ast}]{^{1}_{~1}}\right)&=-\frac{1}{2}\textsl{g}^{00} \frac{d\textsl{g}_{00}}{dr}\left([T^{\ast}]{^{0}_{~0}}-[T^{\ast}]{^{1}_{~1}}\right)
-\textsl{g}^{22} \frac{d \textsl{g}_{22}}{dr}\left([T^{\ast}]{^{1}_{~1}}-[T^{\ast}]{^{2}_{~2}}\right),
\end{align}
where $[T^{\ast}]{^{2}_{~2}}=[T^{\ast}]{^{3}_{~3}}$ due to spherical
symmetry. Therefore, under the GD approach Eq. \eqref{s12} reads
\begin{align}\label{s13}
 \frac{dP_{r}}{dr}&=-\frac{\phi_{1}'}{2}\left(\sigma-P_{r}\right)
-\frac{2}{r}\left(P_{\theta}
-P_{r}\right)+\beta \left[\left(\mathbb{X}^{1}_{~1}\right)'
-\frac{\phi_{1}'}{2}\left(\mathbb{X}^{0}_{~0}-\mathbb{X}^{1}_{~1}\right)
-\frac{2}{r}\left(\mathbb{X}^{2}_{~2}\right)\right],
\end{align}
which is a linear combination of the Eqs. \eqref{s8}-\eqref{s10}.
Additionally, the values of $\sigma$, $P_{r}$, and $P_{\theta}$ correspond to the effective energy density, effective radial pressure, and effective tangential pressure, respectively, which are determined by
\begin{align}\label{s14}
&\sigma^{\ast}=T^{0}_{~0}+\beta\mathbb{X}^{0}_{~0}\equiv\sigma+\sigma^{\mathbb{X}},
\\\label{s15}
&P_{r}^{\ast}=-T^{1}_{~1}-\beta\mathbb{X}^{1}_{~1}\equiv p_{r}+P^{\mathbb{X}}_{r},
\\\label{s16}
&P_{\theta}^{\ast}=-T^{2}_{~2}-\beta\mathbb{X}^{2}_{~2}\equiv p_{\bot}+P^{\mathbb{X}}_{\theta},
\end{align}
which are associated with the SET $[T^{\ast}]_{\mu\nu}$. Then, by making use of the above-stated physical variables, the pressure anisotropy for the source $[T^{\ast}]_{\mu\nu}$ can be defined as
\begin{align}\label{s17}
{\Delta^{\ast}\equiv P^{\ast}_{r}-P^{\ast}_{\theta}= \Delta+\Delta^{\mathbb{X}}.}
\end{align}
Here,
\begin{align}\label{s18}
{\Delta\equiv P_{r}-P_{\theta}, \quad \Delta^{\mathbb{X}}\equiv P^{\mathbb{X}}_{r}-P^{\mathbb{X}}_{\theta}=-\alpha(\mathbb{X}^{1}_{~1}-\mathbb{X}^{2}_{~2}),}
\end{align}
where $\Pi$ and $\Pi^{\mathbb{X}}$ denote the pressure anisotropies corresponding to seed-sector and $\mathbb{X}$-gravitational sector respectively. It is interesting to observe that even if the seed sector is isotropic, the $\mathbb{X}$-gravitational sector will tend to produce anisotropy in the compact system because, in general, $\mathbb{ X}^{1}_{~1}\neq \mathbb{X}^{2}_{~2}$. Now, we will construct the gravitational field equations corresponding to the seed sector as well as the $\theta$-gravitational sector by employing the MGD-decoupling approach.

\section{Field Equations for $[T^{\ast}]_{\mu\nu}$ and $\mathbb{X}_{\mu\nu}$ sectors via MGD-decoupling}

In this section, we will examine how the gravitational system can be
decomposed into two independent gravitational sectors such that the
equations of motion associated with the additional fluid
$\mathbb{X}_{\mu\nu}$ correspond to the effective ``quasi-Einstein
system". We will now apply the GD approach by assuming a solution
for the system {Eqs.} \eqref{s8}--\eqref{s10}, with $\beta=0$, whose
dynamics can be described as the metric ansatz
\begin{align}\label{s19}
ds^{2}=e^{\lambda(r)}dt^{2}-e^{\xi(r)}dr^{2}-r^{2}d\Omega^{2},
\end{align}
with
\begin{align}\label{s20}
e^{\xi(r)}\equiv1-\frac{8\pi}{r}\int^{r}_{0}x^{2}T^{0}_{~0}(x)dx=1-\frac{2m_{s}}{r},
\end{align}
where $m_{s}$ is the mass function associated with the seed gravitational sector. Ovalle \cite{ovalle2017decoupling} proposed that the influence of the $\mathbb{X}$-gravitational sector on the seed SET, $T_{\mu\nu}$ can be determined by introducing the following geometric transformations.
\begin{align}\label{s21a}
&e^{\lambda}\mapsto e^{\phi_{1}}=e^{\lambda+\beta h},
\\\label{s22}
&e^{-\xi}\mapsto e^{-\phi_{2}}=e^{-\xi}+\beta f,
\end{align}
where $f(r)$ and $h(r)$ are the corresponding deformation functions governed by the free parameter $\beta$. The MGD-decoupling requires that
\begin{align}\label{s21}
h(r)\mapsto0;\quad f(r)\mapsto F^{\diamond}(r).
\end{align}
This shows that the temporal metric potential remains the same, while the presence of the $\mathbb{X}$-gravitational source induces a deformation of the radial metric potential \eqref{s22}. Therefore, all the effects stemming form the $\mathbb{X}$-gravitational sector is encoded in the radial metric potential $\textsl{g}_{rr}(r)=e^{\phi_{2}(r)}$. Thus, under the MGD-decoupling the gravitational system \eqref{s8}-\eqref{s10} is divided as follows:
\begin{enumerate}
  \item the standard EFEs associated with seed gravitational sector
  \begin{align}\label{s222}
G^{\mu}_{~\nu}(\lambda,\xi)=8\pi T^{\mu}_{~\nu}, \quad \nabla_{\mu}T^{\mu}_{~\nu}=0,
\end{align}
where
\begin{align}\label{s23}
&G^{0}_{~0}(\lambda,\xi)=\frac{1}{r^{2}}-e^{-\xi}\left(\frac{1}{r^{2}}-\frac{\xi'}{r}\right),
\\\label{s24}
&G^{1}_{~1}(\lambda,\xi)=\frac{1}{r^{2}}-e^{-\xi}\left(\frac{1}{r^{2}}+\frac{\lambda'}{r}\right),
\\\label{s25}
&G^{2}_{~2}(\lambda,\xi)=G^{3}_{~3}(\sigma,\varphi)=\frac{e^{-\xi}}{4}\left(2\frac{\xi'}{r}
-2\frac{\lambda'}{r}+\lambda'\xi'
-\lambda'^{2}-2\lambda''\right),
\end{align}
  \item the gravitational equations corresponding to the $\mathbb{X}$-gravitational sector
  \begin{align}\label{s26}
\beta \mathcal{G}^{\mu}_{~\nu}(\lambda,\xi,F^{\diamond})=8\pi \mathbb{X}^{\mu}_{~\nu}, \quad \nabla_{\mu}\mathbb{X}^{\mu}_{~\nu}=0,
\end{align}
where
\begin{align}\label{s27}
&\mathcal{G}^{0}_{~0}=-\left(\frac{F^{\diamond}}{r^{2}}+\frac{{f^{\ast}}'}{r}\right),
\\\label{s28}
&\mathcal{G}^{1}_{~1}=-F^{\diamond}\left(\frac{1}{r^{2}}+\frac{\lambda'}{r}\right),
\\\label{s29}
&\mathcal{G}^{2}_{~2}=\mathcal{G}^{3}_{~3}=-\frac{F^{\diamond}}{4}\left(2\lambda''+\lambda'^{2}+2\frac{\lambda'}{r}\right)
-\frac{F^{\diamond}}{4}\left(\lambda'+\frac{2}{r}\right).
\end{align}
\end{enumerate}
The hydrostatic equilibrium equations corresponding to the seed SET and $\mathbb{X}$-gravitational sector read
\begin{align}\label{s30}
\frac{dP_{r}}{dr}=-\frac{m_{s}+4\pi r^{3}P_{r}}{(r-2m_{s})}(\sigma+P_{r})+\frac{2}{r}(P_{\theta}-P_{r}),
\end{align}
and
\begin{align}\label{s31a}
\frac{dP^{\mathbb{X}}_{r}}{dr}=-\frac{m_{s}+4\pi r^{3}P^{\ast}_{r}}{(r-2m_{s})}(\sigma^{\mathbb{X}}+P_{r}^{\mathbb{X}})+\frac{2}{r}
(P^{\mathbb{X}}_{\theta}-P^{\mathbb{X}}_{r}),
\end{align}
respectively. Next, the mass functions associated with both gravitational sectors are related through the expression
\begin{align}\label{s31}
m=m_{s}+\beta m_{\mathbb{X}}, \quad \textmd{where} \quad m_{\mathbb{X}}=-\frac{r}{2}F^{\diamond}.
\end{align}
Here, the mass function induced by the $\mathbb{X}$-gravitational sector and the seed SET are denoted by $m_{\mathbb{X}}$ and $m_{s}$, respectively.
Herrera \cite{herrera2018new} proposed that the mass function corresponding to the spherical self-gravitational distribution with anisotropic matter content can be expressed as a sum of two quantities: the homogeneous density and the corresponding alteration due to the non-homogeneous behavior of diversity, which reads
\begin{align}\label{s32}
m=\frac{4\pi}{3}r^{3}\sigma^{\ast}-\frac{4\pi}{3}\int^{r}_{0}\left[\sigma^{\ast}(x)\right]' x^{3}dx.
\end{align}
Then, the corresponding expressions of the mass function for the seed sector and the $\mathbb{X}$-gravitational source take the form
\begin{align}\label{s33}
m=&\frac{4\pi}{3}r^{3}\sigma-\frac{4\pi}{3}\int^{r}_{0}\left[\sigma(s)\right]' s^{3}ds,
\\\label{s34}
m^{\mathbb{X}}=&\frac{4\pi}{3}r^{3}\sigma^{\mathbb{X}}-\frac{4\pi}{3}\int^{r}_{0}\left[\sigma
^{\mathbb{X}}(x)\right]'x^{3}ds.
\end{align}
Furthermore, Tolman's groundbreaking research provided a foundation for understanding the total energy content associated with relativistic self-gravitational objects, which is formally defined as (for details see \cite{tolman1930use})
\begin{align}\label{s35}
m_{T}=4\pi\int^{r_{\Sigma}}_{0}x^{2}e^{(\phi_{1}+\phi_{2})/2}\left([T^{\ast}]^{0}_{~0}-[T^{\ast}]^{1}_{~1}
-2[T^{\ast}]^{2}_{~2}\right)dx,
\end{align}
or, alternatively,
\begin{align}\label{s37}
m_{T}=e^{(\phi_{1}+\phi_{2})/2}\left(m+4\pi r^{3}P^{\ast}_{r}\right).
\end{align}
The above expression can be cast into the form
\begin{align}\label{s38}
m_{T}=\left[m_{T}(r)\right]_{\Sigma}\left(\frac{r}{r_{\Sigma}}\right)^{3}-r^{3}\int^{r_{\Sigma}}_{r}
\left[\frac{4\pi}{x^{4}}\int^{r}_{0}[\sigma^{\ast}(x)]' x^{3}ds-\frac{8\pi}{x}\Delta^{\ast}\right]
e^{(\phi_{1}+\phi_{2})/2}dx,
\end{align}
which can be expressed in terms of Weyl scalar $E$ as
\begin{align}\label{s39}
m_{T}=\left[m_{T}(r)\right]_{\Sigma}\left(\frac{r}{r_{\Sigma}}\right)^{3}+r^{3}\int^{r_{\Sigma}}_{r}
\frac{1}{x}e^{(\phi_{1}+\phi_{2})/2} \left[4\pi\Delta^{\ast}+E\right]dx.
\end{align}
Across the boundary of the compact star, the continuity of the first and second fundamental forms imposes a matching condition for the compact object. This is achieved by ensuring that the interior and exterior metrics, which describe the geometry inside and outside the sphere respectively, match at the boundary surface $r=r_{\Sigma}$. In this respect, the Schwarzschild exterior metric reads
\begin{align}\label{s40}
ds^{2}=\left(1-\frac{2M}{r}\right)dt^{2}-\frac{dr^{2}}{\left(1-\frac{2M}{r}\right)}-r^{2}d\Omega^{2}.
\end{align}
The necessary and sufficient conditions at the boundary can be obtained by matching two metrics and applying the first and second fundamental forms as
\begin{align}\label{s41}
\left(1-\frac{2M}{r}\right)\Big|_{\Sigma}&=e^{\nu}\Big|_{\Sigma},
\\\label{s42}
\left(1-\frac{2M}{r}\right)\Big|_{\Sigma}&=e^{-\lambda}\Big|_{\Sigma},
\\\label{s43}
P^{\ast}_{r}\Big|_{\Sigma}&=0.
\end{align}

\section{Complexity Factor and Other Scalars Under MGD-decoupling}

To explore the inherent features of relativistic compact self-gravitational structures endowed with anisotropic matter configurations, Herrera \cite{herrera2018new} coined the term ``complexity factor". This factor is primarily a scalar function derived from the orthogonal decomposition of the Riemann-Christoffel tensor and is referred to as such because it plays a role in computing the complexity of anisotropic compact stars. Herrera and his coworkers \cite{herrera2009structure} analyzed the growth and structure of spherical anisotropic fluid distributions by constructing a full set of equations in terms of five scalar functions that reduce to $\{{X_{\mathrm{T}}, X_{\mathrm{TF}}, Y_{T}, Y_{\mathrm{TF}}}\}$ in the case of non-dissipative conditions. Additionally, they have demonstrated that all possible solutions to the EFEs corresponding to the time-independent spherical fluid spheres can be expressed explicitly in terms of these scalar quantities. These properties include pressure anisotropy, inhomogeneous density, Tolman mass, heat flux as well as the homogeneous distribution of density. These scalar functions appear as trace and trace-free parts of the specific tensor terms (see \cite{herrera2009structure,herrera2011role} for details)
\begin{align}\label{s44}
Y_{\mu\nu}&=\left(S_{\mu}S_{\nu}-\frac{1}{3}h_{\mu\nu}\right)Y^{\ast}_{\mathrm{TF}}+\frac{1}{3}h_{\mu\nu}
Y^{\ast}_{\mathrm{T}},
\\\label{s45}
X_{\mu\nu}&=\left(S_{\mu}S_{\nu}-\frac{1}{3}h_{\mu\nu}\right)X^{\ast}_{\mathrm{TF}}+\frac{1}{3}h_{\mu\nu}
X^{\ast}_{\mathrm{T}}.
\end{align}
The term $h_{\mu\nu}$ represents the projection tensor. This tensor allows us to project vectors and tensors onto specific surfaces within spacetime, facilitating the analysis and manipulation of geometric structures. The projection tensor enables us to decompose various quantities, such as velocity, momentum, or energy density, into components that are parallel and perpendicular to the hypersurfaces under consideration. The vector $S_\mu$ represents the worldline (path through spacetime) of a massless particle traveling along a null geodesic.
The expressions of the structural scalars in terms of physical variables read
\begin{align}\label{s46}
&Y^{\ast}_{\mathrm{TF}}=4\pi \Delta^{\ast}+E
=8\pi\Delta^{\ast}-\frac{4\pi}{r^{3}}\int^{r}_{0}\left[\sigma^{\ast}(x)\right]'x^{3}dx,
\\\label{s47}
&Y^{\ast}_{\mathrm{T}}=4\pi (\sigma^{\ast}+3P^{\ast}_{r}-2\Delta^{\ast}),
\\\label{s48}
&X^{\ast}_{\mathrm{TF}}=4\pi \Delta^{\ast}-E
=\frac{4\pi}{r^{3}}\int^{r}_{0}\left[\sigma^{\ast}(x)\right]'x^{3}dx,
\\\label{s49}
&X^{\ast}_{\mathrm{T}}=8\pi\sigma^{\ast}(r),
\end{align}
where the values of $\Delta^{\ast}$ and $\Delta$ are defined in Eqs. \eqref{s17} and \eqref{s18}. Then, according to the MGD decoupling, the aforementioned scalars split as
\begin{align}\label{s50}
Y^{\ast}_{\mathrm{TF}}\equiv Y_{\mathrm{TF}}+Y_{\mathrm{TF}}^{\mathbb{X}}; \quad
Y^{\ast}_{\mathrm{T}}\equiv Y_{\mathrm{T}}+Y^{\mathbb{X}}_{\mathrm{T}}; \quad
X^{\ast}_{\mathrm{TF}}\equiv X_{\mathrm{TF}}+X_{\mathrm{TF}}^{\mathbb{X}}; \quad
X^{\ast}_{\mathrm{T}}\equiv X_{\mathrm{T}}+X_{\mathrm{T}}^{\mathbb{X}},
\end{align}
where
\begin{align}\label{s51}
&Y_{\mathrm{TF}}=4\pi \Delta+E=8\pi\Delta-\frac{4\pi}{r^{3}}\int^{r}_{0}\left[\sigma(x)\right]'x^{3}dx,
\\\label{s52}
&Y_{\mathrm{TF}}^{\mathbb{X}}=4\pi \Pi^{\mathbb{X}}+E=8\pi\Delta^{\mathbb{X}}-\frac{4\pi}{r^{3}}\int^{r}_{0}\left[\sigma^{\mathbb{X}}(x)
\right]'x^{3}dx,
\\\label{s53}
&Y_{\mathrm{T}}=4\pi(\sigma+3P_{r}-2\Delta);
\quad
Y^{\mathbb{X}}_{\mathrm{T}}=4\pi(\sigma^{\mathbb{X}}+3P^{\mathbb{X}}_{r}-2\Delta^{\mathbb{X}}),
\\\label{s54}
&X_{\mathrm{TF}}=4\pi \Delta-E=\frac{4\pi}{r^{3}}\int^{r}_{0}\left[\sigma(x)\right]'x^{3}dx; \quad X^{\mathbb{X}}_{\mathrm{TF}}=4\pi \Pi^{\mathbb{X}}-E=\frac{4\pi}{r^{3}}\int^{r}_{0}\left[\sigma^{\mathbb{X}}(x)\right]'x^{3}dx,
\\\label{s55}
&X_{\mathrm{T}}=8\pi\sigma, \quad X^{\mathbb{X}}_{\mathrm{T}}=8\pi\sigma^{\mathbb{X}}.
\end{align}
In this context, the scalars $X^{\ast}_{\mathrm{T}}$ and $X^{\ast}_{\mathrm{TF}}$ denote the density homogeneity and density inhomogeneity of the anisotropic compact star, while the remaining scalars $Y^{\ast}_{\mathrm{T}}$ and $Y^{\ast}_{\mathrm{TF}}$ denote the strong energy condition and the complexity factor, respectively. Next, the local anisotropy in decoupled anisotropic stellar distribution is governed by the interaction between $X^{\ast}_{\mathrm{TF}}$ and $Y^{\ast}_{\mathrm{TF}}$, which can be expressed as
\begin{align}\label{s56}
Y^{\ast}_{\mathrm{TF}}+X^{\ast}_{\mathrm{TF}}\equiv \left[Y_{\mathrm{TF}}+Y_{\mathrm{TF}}^{\mathbb{X}}\right]+\left[X_{\mathrm{TF}}+X_{\mathrm{TF}}
^{\mathbb{X}}\right]=8\pi\left(\Delta+\Delta^{\mathbb{X}}\right).
\end{align}
The definition of Tolman mass in terms of the complexity factor reads
\begin{align}\label{s57a}
m_{T}=\left(\frac{r}{R}\right)^{3}M_{T}+r^{3}\int^{R}_{r}\frac{e^{(\phi_{1}+\phi_{1})/2}}{x}Y^{\ast}_{\mathrm{TF}}dx.
\end{align}
or, alternatively,
\begin{align}\label{s57}
m_{T}=\left(\frac{r}{R}\right)^{3}M_{T}+r^{3}\int^{R}_{r}\frac{e^{(\phi_{1}+\phi_{1})/2}}{x}Y_{\mathrm{TF}}dx
+\beta r^{3}\int^{R}_{r}\frac{e^{(\phi_{1}+\phi_{1})/2}}{x}Y^{\mathbb{X}}_{\mathrm{TF}}dx,
\end{align}
which describes the influence of the $\mathbb{X}$-gravitational sector regulated by decoupling parameter $\beta$. Furthermore, $M_{T}$ represents the total value of the Tolman mass for the self-gravitational compact star with radius $R$. The value of $m_{T}$, using the strong energy condition $Y^{\ast}_{\mathrm{T}}$, can be written as
\begin{align}\label{s58}
m_{T}=\int^{r}_{0}x^{2}e^{(\phi_{1}+\phi_{2})/2}Y_{T}dx+\beta\int^{r}_{0}x^{2}e^{(\phi_{1}+\phi_{2})/2}
Y^{\mathbb{X}}_{T}dx.
\end{align}
Now, for solving both the gravitational sectors, we will adopt different strategies:
\begin{enumerate}
  \item Seed sector $T_{\mu\nu}$: We will use the Embedding Class I condition (Karmarkar condition) to solve this system. This approach utilizes known solutions to constrain the solution space.
  \item $\mathbb{X}$-gravitational sector $\mathbb{X}_{\mu\nu}$: This gravitational will be solved by applying the constraint of zero complexity factor ($\mathrm{Y}^{\ast}_{TF}=0$).
\end{enumerate}

\subsection{Complexity-free Embedding Class I Solutions}

If the spherical symmetry endowed metric associated with the seed SET $T_{\mu\nu}$ satisfies the Karmarkar condition \cite{karmarkar1948gravitational}, then we have
\begin{align}\label{s59}
R_{1010}R_{2323}=R_{1212}R_{3030}+R_{1220}R_{1330}.
\end{align}
The above condition fulfills the Pandey and Sharma condition \cite{pandey1982insufficiency}, i.e., $R_{2323}\neq0$. Then, according to the seed metric ansatz with geometric variables $\{\lambda,\xi\}$, the Karmarkar condition \cite{karmarkar1948gravitational} provides the following differential equation
\begin{align}\label{s60}
\frac{\xi'\lambda'}{1-e^{\xi(r)}}=\xi'\lambda'-2\lambda''-\lambda'^{2}.
\end{align}
Notably, if a metric ansatz follows this condition, it can be embedded within a five-dimensional pseudo-Euclidean space. This kind of metric ansatz is called an embedding Class I metric.
Next, the integration of Eq. \eqref{s60} provides a significant relationship between the geometric variables $\{\lambda,\xi\}$.
\begin{align}\label{s61}
e^{\lambda(r)}=\left(L_{0}+M_{0}\int\sqrt{e^{\xi(r)-1}}\right)^{2} \quad \textmd{or} \quad
e^{\xi(r)}=\left(1+\frac{\lambda'^{2}e^{\lambda(r)}}{4M_{0}}\right),
\end{align}
where $L_{0}$ and $M_{0}$ are integration constants. Now, we will construct the complexity-free compact star models corresponding to both the gravitation sectors. Firstly, we solve the compact system $\{G_{\mu\nu},T_{\mu\nu}\}$ for the seed sector involving the variables $\{\lambda,\xi\}$. Specifying both the metric potentials unlocks the complete set of physical variables ($\sigma$, $P_{r}$, $P_{\bot}$) for the SET $T_{\mu\nu}$. For physically realistic solutions of EFEs, Lake's algorithms \cite{lake2003all} constraint the source variable $\lambda(r)$. In particular, $\lambda(r)$ needs to have a regular minimum at $r=0$ and increase monotonically. This guarantees a time-independent, spherically symmetric solution to the EFEs, with a regular center $r=0$. Therefore, we assume the source function $e^{\lambda(r)}$ as the Durgapal-IV metric ansatz \cite{durgapal1985analytic}
\begin{align}\label{s62}
e^{\lambda(r)}=D(1+\mathcal{A}r^{2})^{4},
\end{align}
where $\mathcal{A}$ and $D$ are the integration constants with units [$length^{-2}$]. The Durgapal-IV metric ansatz remains regular at the center, $r=0$, and consistently increases positively throughout the interior of the self-gravitational compact star, resulting in a realistic model of a stellar distribution. Now, using the Durgapal-IV metric potential, the embedding Class I condition gives \cite{vaidya1982exact} as
\begin{align}\label{s63}
e^{\xi(r)}=1+\mathcal{A}Br^{2}(1+\mathcal{A}r^{2})^{2},
\end{align}
where $B=16\mathcal{A}D/M_{0}$. The metric potential $e^{\xi(r)}$ may be represented as $e^{\xi(r)}=1+O(r^{2})$, which characterizes a necessary condition for the construction of a regular stellar solution at the origin. There, the solution associated with the gravitational system $\{G_{\mu\nu},T_{\mu\nu}\}$ may be expressed through the metric
\begin{align}\label{s64}
ds^{2}=D(1+\mathcal{A}r^{2})^{4}dt^{2}-[1+\mathcal{A}Br^{2}(1+\mathcal{A}r^{2})^{2}]dr^{2}-r^{2}d\Omega^{2}.
\end{align}
We can now obtain the physical variables $\{\sigma,P_{r},P_{\theta}\}$ by inserting the values of the metric potentials $e^{\lambda(r)}$ and $e^{\xi(r)}$ into Eqs. \eqref{s23}-\eqref{s25} as
\begin{align}\label{s65}
&8\pi \sigma=\frac{\mathcal{A}B(1+\mathcal{A}r^{2})[-1+(B-5)\mathcal{A}r^{2}+3B\mathcal{A}^{2}r^{4}
+3B\mathcal{A}^{3}r^{6}+B\mathcal{A}^{4}r^{8}]}{
\left(1+\mathcal{A}Br^{2}+2\mathcal{A}^{2}Br^{4}+\mathcal{A}^{3}Br^{6}\right)^{2}},
\\\label{s66}
&8\pi P_{r}=-\frac{\mathcal{A}[-8+B(1+\mathcal{A}r^{2})^{3}]}{\left(1+\mathcal{A}r^{2}\right)
\left(1+\mathcal{A}Br^{2}+2\mathcal{A}^{2}Br^{4}+\mathcal{A}^{3}Br^{6}\right)},
\\\label{s67}
&8\pi P_{\theta}=\frac{\mathcal{A}[8+16\mathcal{A}r^{2}-B(1+\mathcal{A}r^{2})^{2}(-1+9\mathcal{A}^{2}r^{4})]}
{\left(1+\mathcal{A}r^{2}\right)^{2}
\left(1+\mathcal{A}Br^{2}+2\mathcal{A}^{2}Br^{4}+\mathcal{A}^{3}Br^{6}\right)}.
\end{align}
Now, we solve the gravitational system $\{\mathcal{G}_{\mu\nu},\mathbb{X}_{\mu\nu}\}$ for the $\mathcal{X}$-gravitational sector involving the variables $\{\lambda,\xi,F^{\diamond}\}$ by employing the complexity-free condition, i.e., $Y^{\ast}_{\mathrm{TF}}=0$ with $Y_{\mathrm{TF}}\neq0$. Them, from Eq.\eqref{s46}, we have
\begin{align}\label{s68}
Y^{\ast}_{\mathrm{TF}}=4\pi \Delta^{\ast}+E\equiv Y_{\mathrm{TF}}+Y_{\mathrm{TF}}^{\mathbb{X}}=0,
\end{align}
where $Y_{\mathrm{TF}}$ and $Y_{\mathrm{TF}}^{\mathbb{X}}$ are the specific complexity factors associated with the seed sector and the $\mathbb{X}$-gravitational sector, respectively. The above equation implies
\begin{align}\label{s69}
Y^{\ast}_{\mathrm{TF}}=8\pi\Delta^{\ast}-\frac{4\pi}{r^{3}}\int^{r}_{0}\left[\sigma^{\ast}(x)\right]'x^{3}dx=0.
\end{align}
As density $(\sigma)$, radial pressure $(P_{r})$, and tangential pressure $(P_{\theta})$ are fully determined by Eqs. \eqref{s65}-\eqref{s67}, the expression for $Y_{\mathrm{TF}}$ can be directly obtained as
\begin{align}\label{s70}
Y_{\mathrm{TF}}=\frac{4\mathcal{A}^{2}r^{2}[-2+B(1+\mathcal{A}r^{2})^{3}]}{\left(1+\mathcal{A}r^{2}\right)^{2}
\left(1+\mathcal{A}Br^{2}+2\mathcal{A}^{2}Br^{4}+\mathcal{A}^{3}Br^{6}\right)^{2}}.
\end{align}
Building on Eq. \eqref{s46}, we need to satisfy the following condition
\begin{align}\label{s71}
Y^{\ast}_{\mathrm{TF}}
=8\pi\Delta^{\ast}+8\pi\Delta^{\mathbb{X}}-\frac{4\pi}{r^{3}}\int^{r}_{0}\left[\sigma^{\mathbb{X}}
(x)\right]'x^{3}dx=0,
\end{align}
with $\beta=1$. Next, substituting Eqs. \eqref{s27}\textendash\eqref{s29} into the aforementioned equation, we have
\begin{align}\label{s72}
\lambda'{F^{\diamond}}'+\frac{1}{r}\left[2r\lambda''+\left(r\lambda'-2\right)\lambda'\right]F^{\diamond}
-4Y_{\mathrm{TF}}=0,
\end{align}
whose solution can be obtained by using the Durgapal-IV metric ansatz \eqref{s62} along with Eq. \eqref{s71} as
\begin{align}\label{s73}
F^{\diamond}(r)=\frac{C_{1}}{(1+\mathcal{A}r^{2})^{2}}-\frac{1}{1+\mathcal{A}Br^{2}+2\mathcal{A}^{2}
Br^{4}+\mathcal{A}^{3}Br^{6}},
\end{align}
where $C_{1}$ is an arbitrary integration constant. The specific value of $C_{1}$ can be determined by using a physically relevant condition, which reads
\begin{align}\label{s74}
e^{-\phi_{2}(0)}=e^{-\xi(0)}+\beta F^{\diamond}(0)=1\Rightarrow F^{\diamond}(0)=1.
\end{align}
Therefore, the deformed radial metric potential takes the following form
\begin{align}\label{s76}
e^{-\phi_{2}(r)}=\frac{1+[1+\mathcal{A}Br^{2}(1+\mathcal{A}r^{2})^{2}]\beta F^{\diamond}}{[1+\mathcal{A}Br^{2}(1+\mathcal{A}r^{2})^{2}]},
\end{align}
and the corresponding metric reads
\begin{align}\label{s77}
ds^{2}=D(1+\mathcal{A}r^{2})^{4}dt^{2}-\frac{[1+\mathcal{A}Br^{2}(1+\mathcal{A}r^{2})^{2}] }{1+[1+\mathcal{A}Br^{2}(1+\mathcal{A}r^{2})^{2}]\beta F^{\diamond}}dr^{2}-r^{2}d\Omega^{2}.
\end{align}
The physical variables associated with the above metric are defined in Appendix A as Eqs. \eqref{s78}\textendash\eqref{s85}.

\section{Physical Analysis of the Minimally Deformed Durgapal-IV metric with $Y^{\ast}_{\mathrm{TF}}=0$}

In this section, the physical feasibility of the anisotropically structured Class I Durgapal-IV metric is explored within the framework of the radial metric deformation scheme (MDG decoupling), satisfying the zero-complexity factor condition. It is well-established that physically and mathematically sound modeling of the relativistic stellar distributions must satisfy certain physical constraints. In this work, we analyze the salient features associated with the minimally deformed Durgapal-IV relativistic model under a complexity-free background. The physical properties of these types of anisotropic stellar solutions offer valuable insights into the formation and evolution of self-gravitational compact stars.
We discuss the structural features corresponding to the complexity-free Durgapal-IV metric ansatz filled with anisotropic matter content using the radial metric deformation approach in FIGs. \ref{1f}-\ref{4f}. These profiles depict various properties plotted against the radial variable $r$ with decoupling parameter $\beta\in[0,1]$. The characteristics include:
\begin{itemize}
\item Physical variables: energy density ($\sigma^{\ast}$), radial pressure ($P_{r}^{\ast}$), tangential pressure ($P^{\ast}_{\theta}$) and pressure anisotropy ($\Delta^{\ast}$).

\item Structure scalars: density inhomogeneity ($X^{\ast}_{\mathrm{TF}}$), complexity factor ($Y^{\ast}_{\mathrm{TF}}$), strong energy condition ($Y^{\ast}_{\mathrm{T}}$) and density homogeneity ($X^{\ast}_{\mathrm{T}}$).
\end{itemize}

\subsection{Behavior of Physical Variables}

\begin{figure}[H]
\centering{{\includegraphics[height=2.5 in, width=3.0
in]{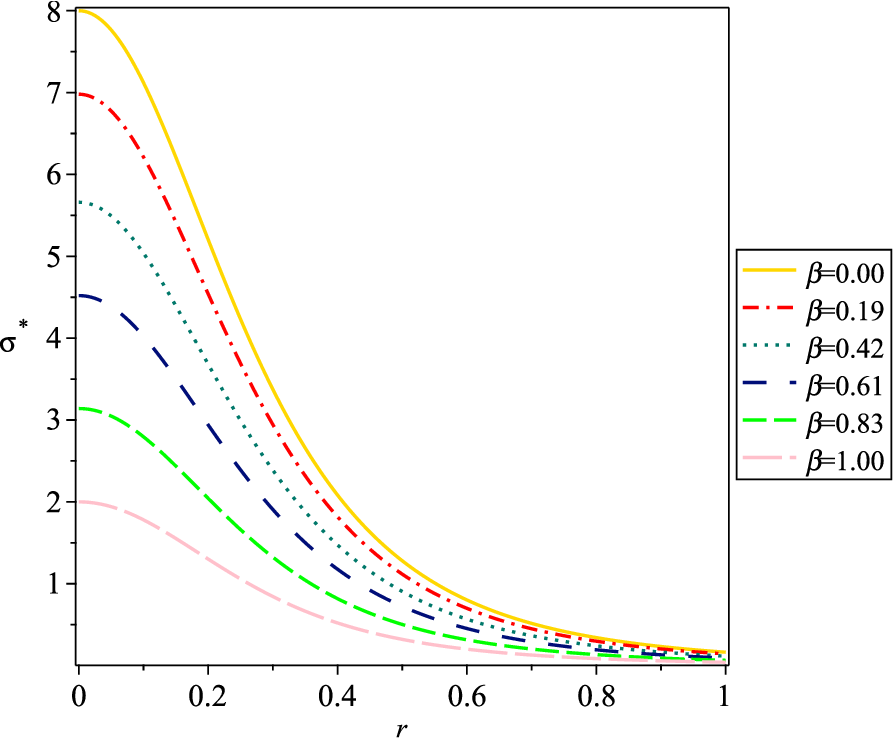}}} \centering{{\includegraphics[height=2.5 in,
width=3.0 in]{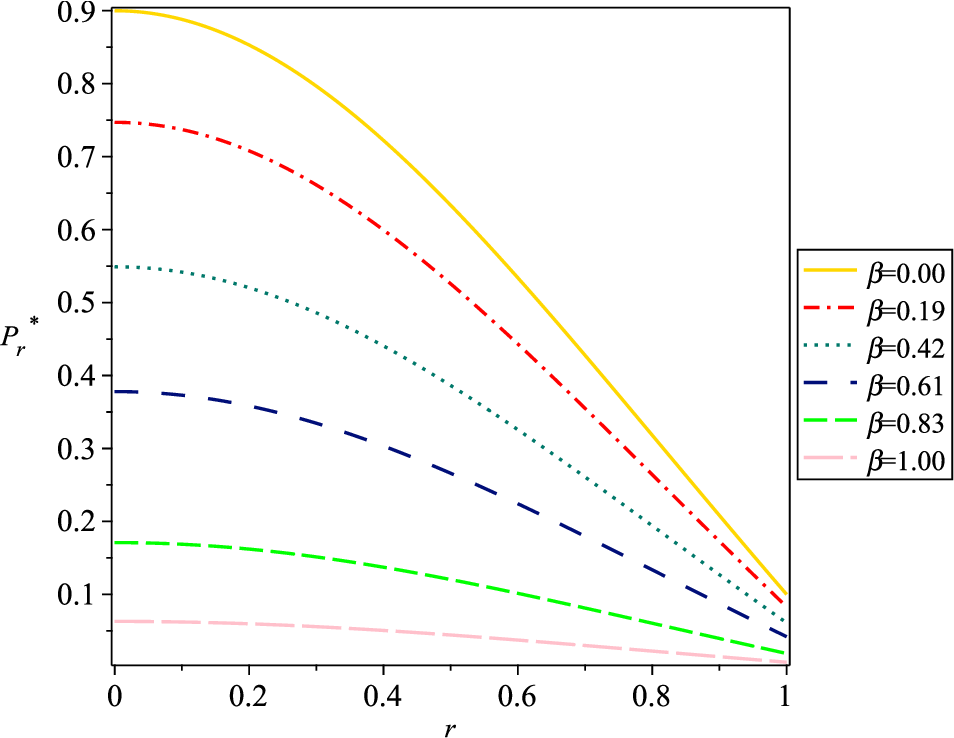}}} \caption{Plot of the energy density
$[\sigma^{\ast}(r,\beta)\times 10^{4}]$ (left panel) and radial
pressure $[P^{\ast}_{r}(r,\beta)\times 10^{4}]$ (right panel) for
the Durgapal-IV model against the radial variable $r$ with different
values of $\beta\in[0,1]$.}\label{1f}
\end{figure}

\begin{figure}[H]
\centering{{\includegraphics[height=2.5 in, width=3.0 in]{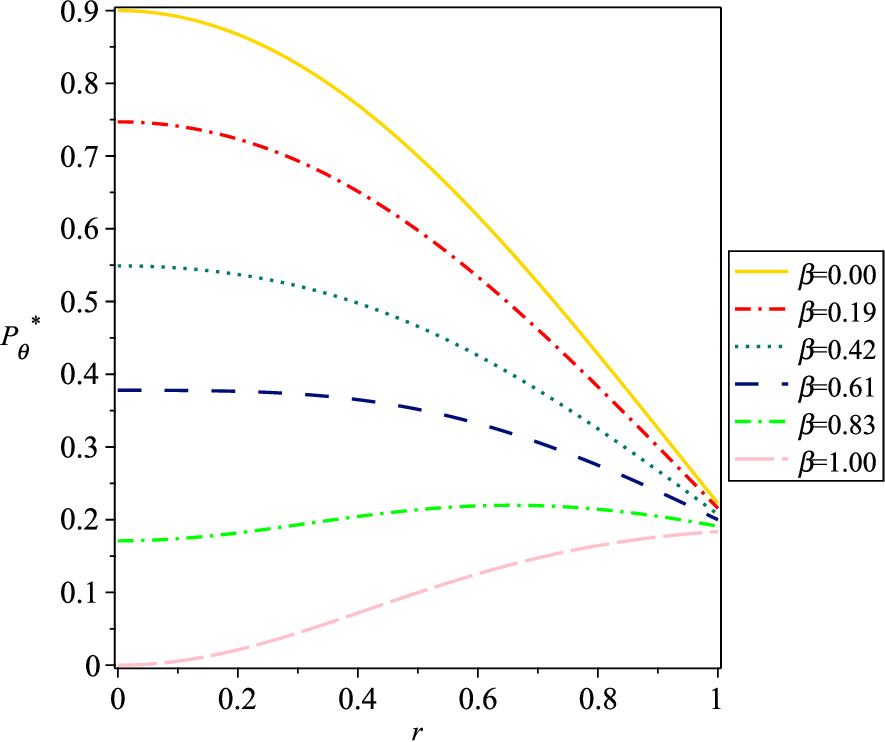}}}
\centering{{\includegraphics[height=2.5 in, width=3.0
in]{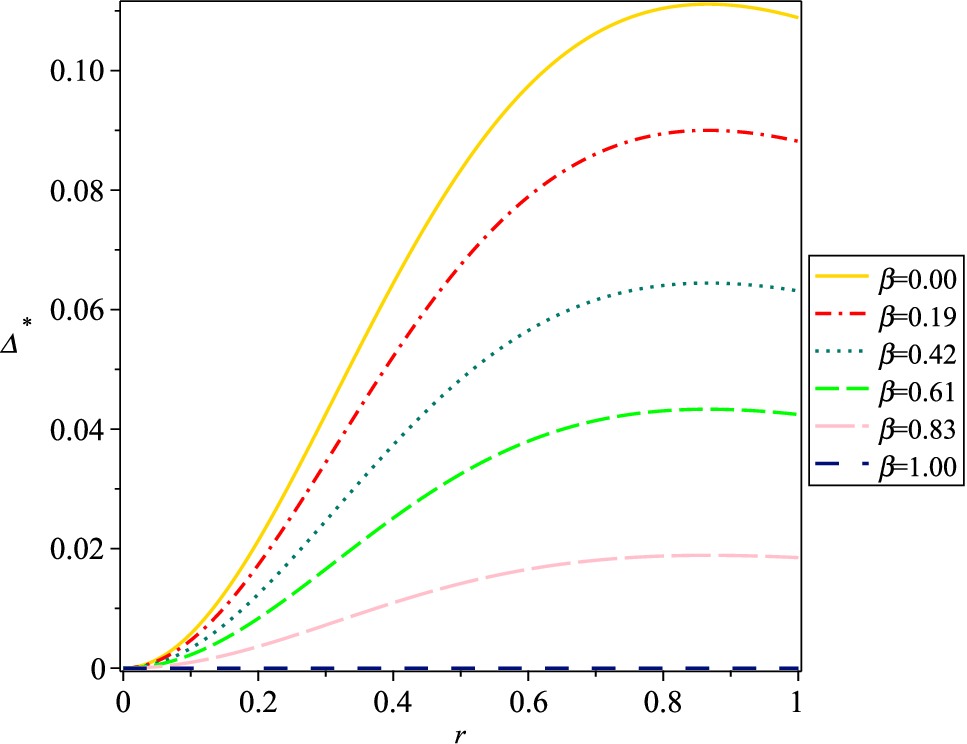}}} \caption{Plot of the tangential pressure
$[P^{\ast}_{\theta}(r,\beta)\times 10^{4}]$ (left panel) and
anisotropic factor $[\Delta^{\ast}(r,\beta)\times 10^{4}]$ (right
panel) corresponding to the Durgapal-IV model versus $r$ with
different $\beta$-values.}\label{2f}
\end{figure}

The behavior of energy density and stress components for different
$\beta$-values are displayed in FIGs. \ref{1f} and \ref{2f}. Their
profiles describe that $\sigma$, $P_{r}$, and $P_{\theta}$ are
gradually decreasing functions towards the stellar's boundary and
exhibit a maximum at the center of the object, as required. We observe that $\sigma^{\ast}$, $P_{r}^{\ast}$, and $P_{\theta}^{\ast}$ decrease monotonically towards the stellar's surface for $\beta=$0.0, 0.19, 0.42, 0.61, 0.83 and 1 as shown in FIGs. \ref{1f} and \ref{2f}.
Similarly, the anisotropy $\Delta^{\ast}$ decreases as we move away from the center for $\beta=$0.0, 0.19, 0.42, 0.61, and 0.83 as displayed in the left panel of FIG. \ref{2f}. However, for $\beta=1$, anisotropy vanishes entirely throughout the model. This disappearance of anisotropy signifies a Class I isotropic solution, i.e., $P_{r}^{\ast}=P_{\theta}^{\ast}$ for all $r\in[0,R]$. This is consistent with a matter distribution described by the stress-energy tensor $T^{\ast}_{\mu\nu}$. The vanishing of pressure anisotropy reveals an important physical insight regarding the
deviation of the anisotropic stellar-type configuration into the
isotropic regime. Because under a zero-complexity factor background,
with no explicit constraints imposing isotropy, the anisotropic
matter content (represented by the SET, $T^{\ast}_{\mu\nu}$)
effectively transforms into an isotropic fluid ($P_{r} =
P_{\theta}$) without any explicit constraints enforcing isotropy.
This interconnection between vanishing complexity and MGD
decoupling offers a novel approach to constructing Durgapal-IV
isotropic stellar models by assuming an initially anisotropic matter
content. This finding has strong physical implications. Because,
generally, GD has been employed for extending isotropic solutions to
anisotropic domains or exploring more complex anisotropic scenarios.
This solution shows that the nature of the matter distribution can
be strongly influenced by the complexity factor in time-independent
spherical stellar configurations. This points to a possible
the connection between the geometric complexity of the system and the
physical characteristics of the filled matter. Further investigation
into this interconnection could provide valuable insights into the
composition of relativistic stellar fluids, shedding light on the
intricate relationships between their geometric and matter
variables.

\subsection{Behavior of Complexity factor}

The variation of $Y^{\ast}_{\mathrm{TF}}$ for the minimally deformed Durgapal-IV solution is displayed in FIG. \ref{3f} (left panel) for different $\beta$-values. It is observed that $Y^{\ast}_{\mathrm{TF}}$ increases initially for a certain distance, followed by a decrease towards the boundary of the compact distribution. The value of $Y^{\ast}_{\mathrm{TF}}$ decreases as the parameter $\beta$ is increased, and it vanishes when $\beta=1$. This suggests that the complexity factor has less of an impact on the gravitationally decoupled stellar distribution. Another important finding is that $Y^{\ast}_{\mathrm{TF}}$ influences the matter variables, such as density and pressure. It affects the surface density, central pressure, and central density because the decline of these matter variables is directly linked with $Y^{\ast}{\mathrm{TF}}$. Furthermore, we see that the complexity factor $Y^{\ast}_{\mathrm{TF}}$ increases as we move towards the boundary of the model as displayed in FIG. \ref{3f} (left panel). However, for higher values of the decoupling constant $\beta$, the overall complexity factor decreases.

\subsection{Energy Condition and Energy Density: Homogeneous and Inhomogeneous Scenarios}

\begin{figure}[H]
\centering{{\includegraphics[height=2.5 in, width=3.0
in]{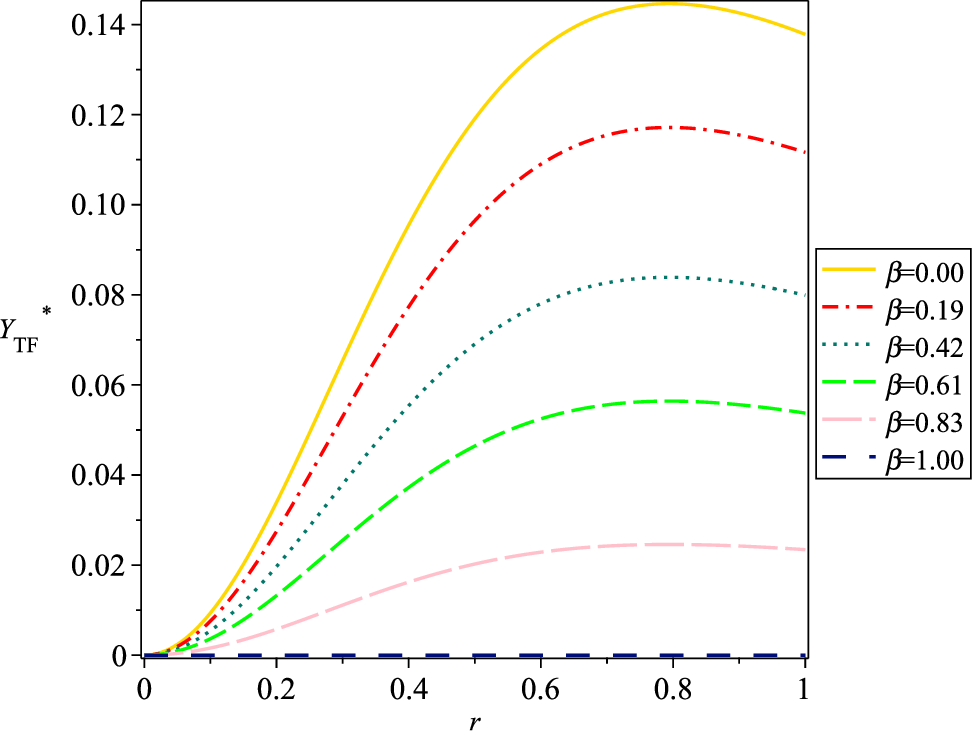}}} \centering{{\includegraphics[height=2.5 in,
width=3.0 in]{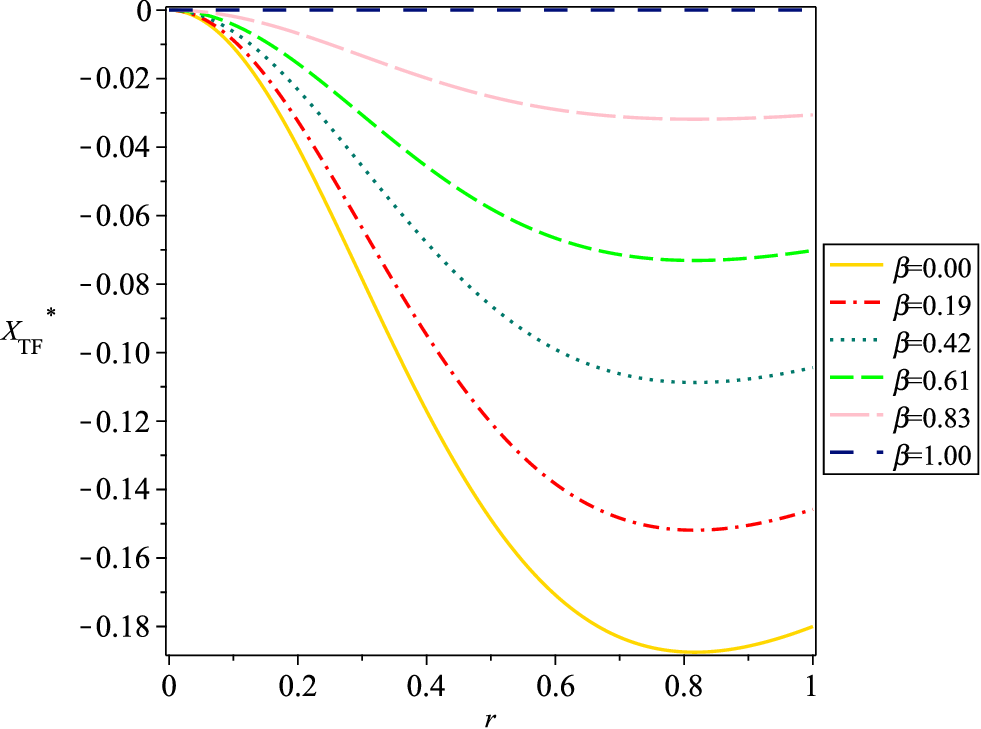}}} \caption{Plot of the complexity factor
$[Y^{\ast}_{\mathrm{TF}}(r,\beta)\times 10^{4}]$ (left panel) and
density inhomogeneity $[X^{\ast}_{\mathrm{TF}}(r,\beta)\times
10^{4}]$ (right panel) subject to the Durgapal-IV model versus $r$
with different values of the coupling constant $\beta$.}\label{3f}
\end{figure}

\begin{figure}[H]
\centering{{\includegraphics[height=2.5 in, width=3.0 in]{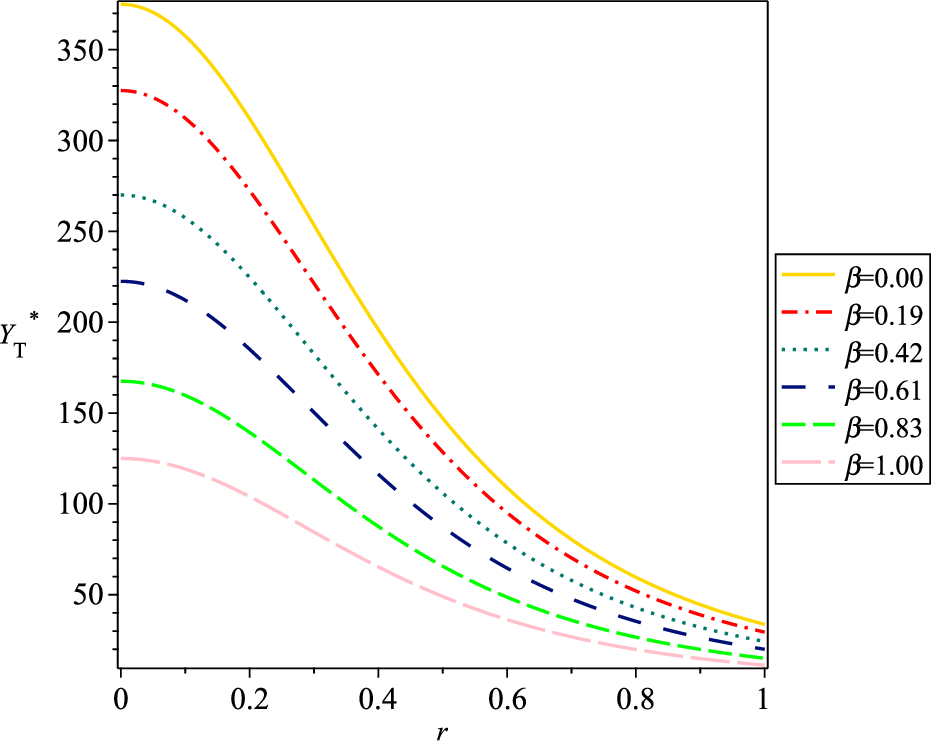}}}
\centering{{\includegraphics[height=2.5 in, width=3.0 in]{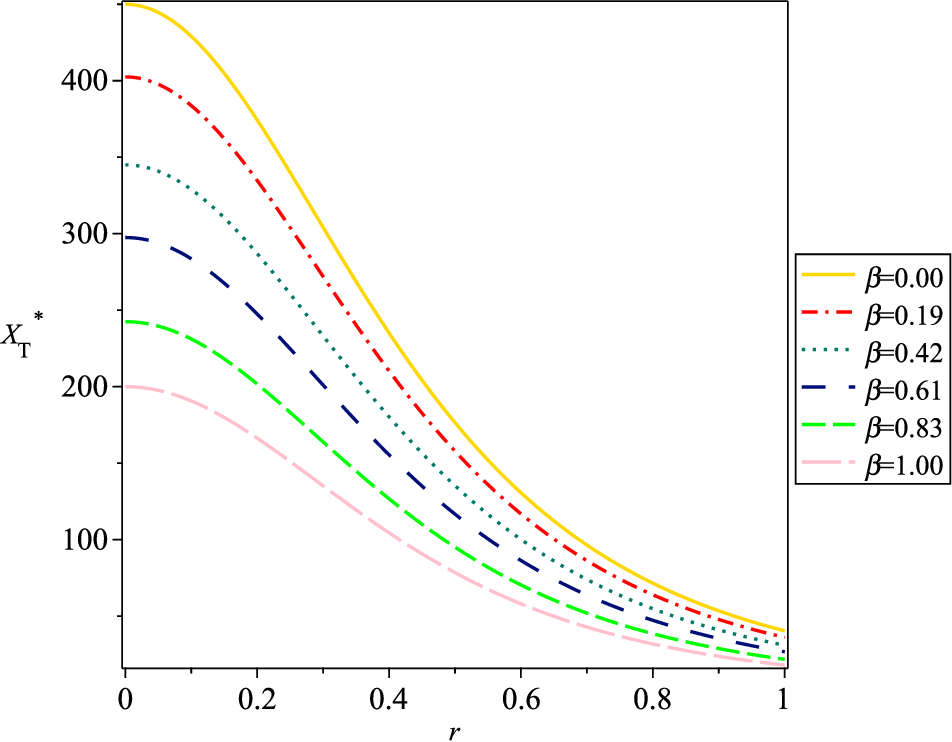}}}
\caption{Plot of strong energy condition
$[Y^{\ast}_{\mathrm{T}}(r,\beta)\times 10^{4}]$ (left panel) and
density homogeneity $[X^{\ast}_{\mathrm{T}}(r,\beta)\times 10^{4}]$
(right panel) for the Durgapal-IV model versus $r$ with
$\beta\in[0,1]$.}\label{4f}
\end{figure}

The behavior of other structural scalars, including the density homogeneity, strong energy condition, and density inhomogeneity, is observed for the permissible values of the parameter $\beta$ as displayed in FIGs. \ref{3f} and \ref{4f}. The solution profile encompasses the entire parameter space of the decoupling constant $\beta$. This profile satisfies all relevant scalar constraints, ensuring its physical viability. These constraints include the complexity factor, density inhomogeneity, strong energy conditions, and homogeneous energy density distribution. On the other hand, it is observed that as the value of $\beta$ increases from 0 to 1, the values of $Y^{\ast}_{T}$ and $X^{\ast}_{T}$ decrease as shown in FIG. \ref{4f} Furthermore, the profile of $X^{\ast}_{\mathrm{TF}}$ displays that its value increases as $\beta$ increases from 0 to 1 as shown in FIG. \ref{4f} (right panel).

\section{Conclusions}

Determining the presence of anisotropic matter distribution in compact stars through direct observation is currently a significant challenge. The internal structure of compact stars cannot be directly observed using current observational methods. However, observations of phenomena like gravitational waves or extreme matter behavior can provide indirect clues about the internal structure of compact stars. Regarding the transition from an anisotropic matter configuration into an isotropic domain within compact stars, theoretical possibilities exist, but the extent to which this process occurs in the majority of compact stars remains uncertain. Evolutionary models and simulations indicate that physical processes like accretion or nuclear reactions can cause changes in the matter distribution within compact stars over time. However, the prevalence of such transformations among compact stars is an active area of research and debate.
The MGD decoupling approach offers a straightforward and efficient
framework for generating well-behaved solutions for EFEs. Despite its
apparent simplicity, this approach provides profound insights into
self-gravitating anisotropic configurations. Its primary advantage
lies in breaking down a complex gravitational system into two
simpler, decoupled sets. The first set corresponds to the standard
EFEs for the considered seed matter content, while the second,
influenced by the additional gravitational source term
$\mathbb{X}{\mu\nu}$, governs the anisotropic sector. This second
system is often termed the quasi-Einstein gravitational system,
leading to a departure from the standard EFEs. In this context, we have considered the
Durgapal-IV model for exploring the dynamics of complexity-free stellar distribution.
Subsequently, the quasi-Einstein equations necessitate supplementary
information, either through constraints such as an EoS or the
physically acceptable condition, such as a zero-complexity factor,
for the source term $\mathbb{X}{\mu\nu}$ or by specifying an
appropriate representation of the MGD-function $F^{\diamond}(r)$, as
implemented in this study. This work demonstrates that the presented
complexity-free Durgapal-IV model of anisotropic fluid spheres
satisfies all necessary criteria for a valid solution. In this work,
we have merged the embedding Class I approach with gravitational
decoupling via the MGD scheme to construct complexity-free anisotropic
solutions for stellar interiors. To investigate the potentiality of
finding simple, analytical solutions for non-rotating, spherically
symmetric, and stationary stellar structures with zero-complexity
factor $(Y^{\ast}_{\mathrm{TF}}=0)$, we use the MGD approach
combined with the Durgapal-IV metric model. In this direction, we
utilize the combination of the Class I condition and the condition
$Y^{\ast}_{\mathrm{TF}}=0$ as the basis for the MGD function to
construct physically realistic compact star models.

We have analyzed the structural features of self-gravitating,
anisotropic stars within the Durgapal-IV model using gravitational
decoupling. We analyzed various matter variables
$\{\sigma^{\ast},P^{\ast}_{r}, P^{\ast}_{\theta},\Delta^{\ast}\}$
and four scalar quantities
$\{Y^{\ast}_{\mathrm{TF}},X^{}_{\mathrm{TF}},T^{\ast}_{\mathrm{T}},X^{\ast}_{\mathrm{T}}\}$
for $\beta\in[0,1]$. We have noted the impact of the decoupling
parameter $\beta$ on all matter variables within the framework of
the Durgapal-IV relativistic model. The scalar function
$Y^{\ast}_{\mathrm{TF}}$ used for measuring complexity notably
affects the inherent structural characteristics of stellar interiors
and facilitates the transition from an anisotropic to an isotropic
solution. Specifically, when $\beta=1$, indicating a vanishing
complexity factor, solutions derived from the Durgapal-IV ansatz
shift from the initially considered anisotropic regime to the
isotropic domain. This emphasizes the relevance of the complexity
factor on physical features associated with self-gravitational
stellar distributions through the mechanism of GD. In particular, we
probed the captivating structural properties corresponding to the
Durgapal-IV model: the disappearance of the anisotropic factor as
the complexity factor approaches zero ($\beta=1$). This finding
involves the conversion of an anisotropic matter content of stress
tensor constitution $T^{\ast}_{\mu\nu}$ into a homogeneous matter
configuration, where $P_{r}=P_{\theta}$. This transformation is
encouraged by GD, coupled with the condition
$Y^{\ast}_{\mathrm{TF}}=0$, without requiring any isotropic
precondition. Furthermore, we have observed significant trends in
stress components throughout the solution employing the Durgapal-IV
metric potential. Specifically, it has been observed that as
$Y^{\ast}_{\mathrm{TF}}\rightarrow0$, there is a noticeable rise in
pressure throughout the interior of the star, a situation considered
undesirable. Moreover, we have revealed that contributions from GD,
enforced by the constraint $T^{\ast}_{\mathrm{TF}}=0$, act to reduce
the anisotropic factor. Interestingly, incorporating the structure
scalar $T^{\ast}_{\mathrm{TF}}$ can lead to required outcomes,
exerting substantial influence on all matter variables and producing
a feasible solution for the minimum values of $\beta$.

In summary, this research effectively examined the influence of complexity on self-gravitational, anisotropic stars employing the
MGD-based GD approach. This potent method allows for the generation
of novel, physically plausible isotropic models for anisotropic
fluid configurations by assuming the Durgapal metric potential
within the formalism of embedding Class I method. Our results
illustrate the pivotal role of the complexity factor in altering the
distribution of matter within static, self-gravitational stellar
interiors. The influence of complexity on the structural features of anisotropic stellar objects can be explored through the mechanism of the complete geometric deformation approach. However, this study specifically focuses on investigating the complexity of minimally deformed compact systems. The examination of the complexity factor for completely deformed stellar structures could be considered for future investigation. We have provided insights into the geometric properties and behavior of spacetime through matter-geometric relationships.

This finding is significant because the disappearance of pressure anisotropy throughout the model suggests that an anisotropic matter distribution transforms into an isotropic matter content under the paradigm of a zero-complexity factor, without necessitating any isotropy requirement. This result illustrates how the coupling of MGD decoupling with the concept of a zero-complexity factor can be employed to model the isotropic stellar structures that are originally filled with anisotropic matter content. Until now, gravitational decoupling has been utilized to extend isotropic solutions into anisotropic domains or from anisotropic to more potent anisotropic domains. This work differs from previous studies in which the notion of GD is generally used to extend isotropic stellar solutions to the anisotropic domain. In the past, several attempts have been made to extend isotropic stellar configurations into the anisotropic regime within the GD approach, including the Durgapal-IV metric \cite{contreras2022uncharged,andrade2021stellar,das2022anisotropic,maurya2022simple}.
However, this work features a novel aspect of the GD approach: the coupling of GD with a zero-complexity factor may be used to transform the originally anisotropic matter distribution into an isotropic domain with any pre-assumed isotropic condition.

\section*{Appendix A}

The physical variables associated with the $\mathbb{X}$-gravitational sector take the form
\begin{align}\label{s78}
8\pi\sigma^{\mathbb{X}}=&\frac{\beta\mathcal{A}[6+3\mathcal{A}r^{2}+\mathcal{A}^{2}r^{4}+\mathcal{A}B^{2}r^{2}
(1+\mathcal{A}r^{2})^{4}(-1+3\mathcal{A}r^{2})
-B(1+\mathcal{A}r^{2})^{2}(3+\mathcal{A}r^{2}+11\mathcal{A}^{2}r^{4}+5\mathcal{A}^{3}r^{6})]}
{\left(1+\mathcal{A}r^{2}\right)^{3}
\left(1+\mathcal{A}Br^{2}+2\mathcal{A}^{2}Br^{4}+\mathcal{A}^{3}Br^{6}\right)^{2}},
\\\label{s79}
8\pi P_{r}^{\mathbb{X}}=&\frac{\beta\mathcal{A}(1+9\mathcal{A}r^{2})[-2-\mathcal{A}r^{2}+B(1+\mathcal{A}r^{2})^{2}]}{\left(1+\mathcal{A}r^{2}\right)^{3}
\left(1+\mathcal{A}Br^{2}+2\mathcal{A}^{2}Br^{4}+\mathcal{A}^{3}Br^{6}\right)},
\\\nonumber
8\pi P_{\theta}^{\mathbb{X}}=&-\frac{\beta \mathcal{A}[2(-1+3\mathcal{A}r^{2}+20\mathcal{A}^{2}r^{4}+8\mathcal{A}^{3}r^{6})+
B(1+\mathcal{A}r^{2})^{2}(1-2\mathcal{A}r^{2}+12\mathcal{A}^{2}r^{4}+78\mathcal{A}^{3}r^{6}+31\mathcal{A}^
{4}r^{8})]}{\left(1+\mathcal{A}r^{2}\right)^{4}
\left(1+\mathcal{A}Br^{2}+2\mathcal{A}^{2}Br^{4}+\mathcal{A}^{3}Br^{6}\right)^{2}}
\\\label{s80}
&+\frac{\beta \mathcal{A}[2\mathcal{A}^{2}B^{2}(5+13\mathcal{A}r^{2})(r+\mathcal{A}r^{3})^{4}]}
{\left(1+\mathcal{A}r^{2}\right)^{4}
\left(1+\mathcal{A}Br^{2}+2\mathcal{A}^{2}Br^{4}+\mathcal{A}^{3}Br^{6}\right)^{2}},
\end{align}
where $\sigma^{\mathbb{X}}$, $P^{\mathbb{X}}_{r}$, and $P^{\mathbb{X}}_{\theta}$ denote the physical variables associated with the decoupling fluid $\mathbb{X}_{\mu\nu}$.
The anisotropic factor associated with the Durgapal-IV metric potential takes the form
\begin{align}\nonumber
\Delta^{\ast}(r,\beta)=&[\mathcal{A}(-4\beta+\mathcal{A}B^{2}r^{2}(1+\mathcal{A}r^{2})^{4}
(-2-6\mathcal{A}r^{2}+14\mathcal{A}^{3}r^{6}+8\mathcal{A}^{4}r^{8}+\mathcal{A}^{2}r^{4}(2-17\beta)+\beta)
+4(-4+3\beta)
\\\nonumber
&\times\mathcal{A}^{2}r^{4}+\mathcal{A}^{3}r^{6}(-8+7\beta)-\mathcal{A}r^{2}(8+15\beta)+2B(1+\mathcal{A}
r^{2})^{2}(-1-3\mathcal{A}^{2}r^{4}+3
\mathcal{A}r^{2}(\beta-1)+\beta+11\beta \mathcal{A}^{4}
\\\label{s81}
&\times r^{8}+\mathcal{A}^{3}(-1+25\beta)r^{6}))]/
[(1+Ar^{2})^{4}(1+\mathcal{A}Br^{2}+2\mathcal{A}^{2}Br^{4}+\mathcal{A}^{3}Br^{6})^{2}].
\end{align}
The corresponding value of the complexity factor $Y^{\ast}_{\mathrm{TF}}$ turn out to be
\begin{align}\label{s82}
Y^{\ast}_{\mathrm{TF}}(r,\beta)&=(1-\beta)\frac{4A^{2}r^{2}[-2+B(1+Ar^{2})^{3}]}
{[(1+Ar^{2})^{2}(1+\mathcal{A}Br^{2}+2\mathcal{A}^{2}Br^{4}+\mathcal{A}^{3}Br^{6})^{2}]}.
\end{align}
This expression captures the effects of the $\mathbb{X}$-gravitational source on the complexity of self-gravitational compact star by interpolating between the original value of $\beta$, i.e., $\beta=0$  and complexity-free condition ($\beta=1$). The corresponding values of the other structure scalars read
\begin{align}\nonumber
X^{\ast}_{\mathrm{TF}}(r,\beta)=&-\mathcal{A}^{2}r^{2}[B^{2}(1+\mathcal{A}r^{2})^{4}(1
+3\mathcal{A}^{2}r^{4}+\mathcal{A}^{3}r^{6}
-3\mathcal{A}r^{2}(\beta-1)-\beta+2B(1+\mathcal{A}r^{2})^{2}
\\\label{s83}
&\times(\beta-1+\mathcal{A}^{2}r^{4}(2\beta-1)+\mathcal{A}r^{2}(5\beta-2))]/
[(1+Ar^{2})^{3}(1+\mathcal{A}Br^{2}+2\mathcal{A}^{2}Br^{4}+\mathcal{A}^{3}Br^{6})^{2}],
\\\nonumber
Y^{\ast}_{\mathrm{T}}(r,\beta)=&-[\mathcal{A}(-20+70\mathcal{A}^{8}B^{2}r^{16}+11\mathcal{A}^{9}B^{2}
r^{18}+2\mathcal{A}^{7}B^{2}r^{14}(95-16\beta)+2(B-2)\beta+\mathcal{A}r^{2}(-68-20B+B^{2}
\\\nonumber
&+12\beta)+2\mathcal{A}^{6}Br^{12}(-10+B(143-72\beta)+19\beta)+2\mathcal{A}^{4}Br^{8}(-100+B(73-112\beta)
+131\beta)-2\mathcal{A}^{2}r^{4}
\\\nonumber
&\times(38+B(50-9\beta)-26\beta+B^{2}(-5+8\beta))+2\mathcal{A}^{3}r^{6}(B^{2}(25-48\beta)+2(-7+5\beta)+4B(-25+
18\beta))
\\\label{s84}
&-4\mathcal{A}^{5}Br^{10}(25-44\beta+B(-65+64\beta)))]/[(1+\mathcal{A}r^{2})^{3}(1+\mathcal{A}Br^{2}+2\mathcal{A}
^{2}Br^{4}+\mathcal{A}^{3}Br^{6})^{2}],
\\\nonumber
X^{\ast}_{\mathrm{T}}(r,\beta)=&[\mathcal{A}(-4\beta+\mathcal{A}B^{2}r^{2}(1+\mathcal{A}r^{2})^{4}(-2-
6\mathcal{A}r^{2}+14\mathcal{A}^{3}r^{6}
+8\mathcal{A}^{4}r^{8}+\mathcal{A}^{2}r^{4}(2-17\beta)+\beta)+(-4+3\beta)
\\\nonumber
&\times4\mathcal{A}^{2}r^{4}+\mathcal{A}^{3}r^{6}(-8+7\beta)-\mathcal{A}r^{2}(8+15\beta)+
2B(1+\mathcal{A}r^{2})^{2}(-1-3\mathcal{A}^{2}r^{4}+3\mathcal{A}r^{2}(\beta-1)+\beta+11\mathcal{A}^{4}
\\\label{s85}
&\times\beta r^{8}+A^{3}r^{6}(-1+25\beta)))]/[(1+\mathcal{A}r^{2})^{4}(1+\mathcal{A}Br^{2}+2\mathcal{A}
^{2}Br^{4}+\mathcal{A}^{3}Br^{6})^{2}].
\end{align}
where $X^{\ast}_{TF}$, $Y^{\ast}_{TF}$, $X^{\ast}_{T}$, and $Y^{\ast}_{T}$ denote density inhomogeneity, complexity factor, density homogeneity, and the strong energy condition for the compact system, respectively.
\vspace{0.3cm}

\vspace{0.3cm}
\section*{Acknowledgement}

The work of KB was partially supported by the JSPS KAKENHI Grant
Number 21K03547 and 23KF0008. The work by BA was supported by Researchers Supporting Project number: RSPD2024R650, King Saud University, Riyadh, Saudi Arabia.

\section*{Conflict of Interest}

The authors declare no conflict of interest.

\section*{Data Availability Statement}

This manuscript has no associated data
or the data will not be deposited. [Authors comment: This manuscript
contains no associated data.]

\vspace{0.3cm}

\renewcommand{\theequation}{A\arabic{equation}}
\setcounter{equation}{0}

\end{document}